\documentclass[twocolumn,showpacs,preprintnumbers,amsmath,amssymb,nofootinbib]{revtex4-1}
\usepackage{graphicx}
\usepackage{mathtools}
\usepackage{dcolumn}
\usepackage{bm}
\begin{document}

\title{Scale invariance of electrodynamics in radio-frequency linear accelerators}
\author{Osamu Kamigaito}
\email{kamigait@riken.jp}
\affiliation{RIKEN Nishina Center for Accelerator-Based Science\\
2-1 Hirosawa, Wako-shi, Saitama 351-0198, Japan}%

\begin{abstract}
This study discusses the scale transformation of electromagnetic fields and trajectories of ions in radio-frequency linear accelerators.
We will obtain a condition for mechanical similarity, where Maxwell equations with source terms and the equation of motion under Lorentz force 
remain invariant, including space charge fields, while the trajectories of ions are accordingly scaled.
The possibility of linear accelerators for extremely high current beams will be considered based on this condition.

\end{abstract}


\maketitle

\section{\label{sec:scalemot}Scale transformation of the equation of motion}
We will consider the scale transformation of non-relativistic electrodynamic equations.
Suppose that an ion with mass $m$ and charge $q$ is in motion under 
electromagnetic fields $\bm{E}\left(\bm{r}, t\right)$ and $\bm{B}\left(\bm{r}, t\right)$.
In this case, the coordinate function $\bm{x}(t)$ of the ion satisfies the following equation of motion:
\begin{eqnarray}
m\ddot{\bm{x}}(t)=q\left\{\bm{E}\left(\bm{x}(t), t\right)+\dot{\bm{x}}(t)\times\bm{B}\left(\bm{x}(t), t\right)\right\}.
\label{eq:lorentz}
\end{eqnarray}
On the other hand, consider the scale transformation of the aforementioned motion, i.e.,
\begin{eqnarray}
\bm{X}(t)\coloneqq \lambda\bm{x}(\tau t),
\label{eq:scaled}
\end{eqnarray}
where $\lambda$ and $\tau$ are dimensionless positive constants.
What kind of electromagnetic fields would provide $\bm{X}(t)$ as a solution to the equation of motion?
To derive this, we calculate the second-order time derivative of $\bm{X}(t)$ using eq. (\ref{eq:lorentz}), and we obtain
\begin{eqnarray}
m\ddot{\bm{X}}(t)&=&m\lambda \tau^2\ddot{\bm{x}}(\tau t)\nonumber \\
&=&q\lambda \tau^2\left\{\bm{E}\left(\bm{x}(\tau t), \tau t\right)+\dot{\bm{x}}(\tau t)\times\bm{B}\left(\bm{x}(\tau t), \tau t\right)\right\}\nonumber \\
&=&q\lambda \tau^2\left\{\bm{E}\left(\frac{\bm{X}(t)}{\lambda}, \tau t\right)\right.\nonumber \\
&+&\left. \frac{\dot{\bm{X}}(t)}{\lambda \tau}\times\bm{B}\left(\frac{\bm{X}(t)}{\lambda}, \tau t\right)\right\}\nonumber \\
&=&q\left\{\lambda \tau^2\bm{E}\left(\frac{\bm{X}(t)}{\lambda}, \tau t\right)\right.\nonumber \\
&+&\left. \dot{\bm{X}}(t)\times \tau\bm{B}\left(\frac{\bm{X}(t)}{\lambda}, \tau t\right)\right\}.
\end{eqnarray}
Therefore, when we define new electromagnetic fields $\tilde{\bm{E}}\left(\bm{r}, t\right)$ and $\tilde{\bm{B}}\left(\bm{r}, t\right)$
as
\begin{eqnarray}
\tilde{\bm{E}}\left(\bm{r}, t\right)\coloneqq \lambda \tau^2\bm{E}\left(\frac{\bm{r}}{\lambda}, \tau t\right),
\label{eq:enew}
\end{eqnarray}
and
\begin{eqnarray}
\tilde{\bm{B}}\left(\bm{r}, t\right)\coloneqq \tau\bm{B}\left(\frac{\bm{r}}{\lambda}, \tau t\right),
\label{eq:bnew}
\end{eqnarray}
the function $\bm{X}(t)$ satisfies the equation of motion under these new electromagnetic fields:
\begin{eqnarray}
m\ddot{\bm{X}}(t)=q\{\tilde{\bm{E}}\left(\bm{X}(t), t\right)+
\dot{\bm{X}}(t)\times\tilde{\bm{B}}\left(\bm{X}(t), t\right)\}.
\end{eqnarray}

Let us analyze the scale transformation mentioned above in a radio-frequency (rf) linear accelerator (linac).
If we set $\lambda=1$, then the question would be, 
``Can ions be accelerated in the same trajectory in a resonator whose frequency is $\tau$ times larger?''
The answer to this question is that if we increase the electric field by a factor of $\tau^2$ 
and the magnetic field by a factor of $\tau$, acceleration is possible in the same trajectory. 
The speed of the ions will be $\tau$ times faster.
These considerations are necessary when designing a variable-frequency linac\cite{bib:ode84}.

\section{\label{sec:scaleem}Scale transformation of the electromagnetic field}

Next, we consider whether the new electromagnetic fields (\ref{eq:enew}) and (\ref{eq:bnew}) satisfy Maxwell's equations
when the original electromagnetic fields $\bm{E}\left(\bm{r}, t\right)$ and $\bm{B}\left(\bm{r}, t\right)$ do.

Let us assume that $\bm{E}\left(\bm{r}, t\right)$ and $\bm{B}\left(\bm{r}, t\right)$ satisfy Faraday's law, i.e.,
\begin{eqnarray}
\nabla\times\bm{E}(\bm{r}, t)=-\frac{\partial \bm{B}}{\partial t}(\bm{r}, t).
\label{eq:faraday}
\end{eqnarray}
Using eqs. (\ref{eq:enew}) and (\ref{eq:faraday}), we obtain
\begin{eqnarray}
\nabla\times\tilde{\bm{E}}(\bm{r}, t)&=&\tau^2\nabla\times\bm{E}\left(\frac{\bm{r}}{\lambda}, \tau t\right)\nonumber\\
&=&-\tau^2\frac{\partial \bm{B}}{\partial t}\left(\frac{\bm{r}}{\lambda}, \tau t\right).
\end{eqnarray}
On the other hand, based on eq. (\ref{eq:bnew}), we obtain
\begin{eqnarray}
-\frac{\partial \tilde{\bm{B}}}{\partial t}(\bm{r}, t)=-\tau^2\frac{\partial \bm{B}}{\partial t}\left(\frac{\bm{r}}{\lambda}, \tau t\right),
\end{eqnarray}
which implies that (\ref{eq:enew}) and (\ref{eq:bnew}) also satisfy Faraday's law, i.e.,
\begin{eqnarray}
\nabla\times\tilde{\bm{E}}(\bm{r}, t)=-\frac{\partial \tilde{\bm{B}}}{\partial t}(\bm{r}, t).
\end{eqnarray}

Next, we examine Amp\`ere's law in vacuum.
Let us assume that
\begin{eqnarray}
\frac{1}{\mu_0}\nabla\times\bm{B}(\bm{r}, t)=\epsilon_0 \frac{\partial \bm{E}}{\partial t}(\bm{r}, t).
\label{eq:ampere}
\end{eqnarray}
Based on eqs. (\ref{eq:bnew}) and (\ref{eq:ampere}), we obtain
\begin{eqnarray}
\frac{1}{\mu_0}\nabla\times\tilde{\bm{B}}(\bm{r}, t)&=&\frac{\tau}{\lambda\mu_0}\nabla\times\bm{B}\left(\frac{\bm{r}}{\lambda}, \tau t\right)\nonumber\\
&=&\frac{\tau\epsilon_0}{\lambda}\frac{\partial \bm{E}}{\partial t}\left(\frac{\bm{r}}{\lambda}, \tau t\right).
\end{eqnarray}
On the other hand, equation (\ref{eq:enew}) leads to
\begin{eqnarray}
\epsilon_0 \frac{\partial \tilde{\bm{E}}}{\partial t}(\bm{r}, t)=
-\lambda \tau^3\epsilon_0\frac{\partial \bm{E}}{\partial t}\left(\frac{\bm{r}}{\lambda}, \tau t\right).
\end{eqnarray}
When we compare the right-hand sides of the two equations mentioned above, 
we can see that, only under a constraint,
\begin{eqnarray}
\label{eq:kousoku}
\lambda \tau =1,
\end{eqnarray}
the fields (\ref{eq:enew}) and (\ref{eq:bnew}) satisfy Amp\`ere's law in vacuum; i.e.,
\begin{eqnarray}
\frac{1}{\mu_0}\nabla\times\tilde{\bm{B}}(\bm{r}, t)=\epsilon_0 \frac{\partial \tilde{\bm{E}}}{\partial t}(\bm{r}, t).
\end{eqnarray}

Therefore, a scale transformation that satisfies the equation of motion and source-free Maxwell's equations
simultaneously is given as
\begin{eqnarray}
\label{eq:scalex}
\bm{X}(t)&\coloneqq&\lambda\bm{x}(\frac{t}{\lambda}),\\
\label{eq:scalee}
\tilde{\bm{E}}\left(\bm{r}, t\right)&\coloneqq& \frac{1}{\lambda}\bm{E}\left(\frac{\bm{r}}{\lambda}, \frac{t}{\lambda}\right),\\
\label{eq:scaleb}
\tilde{\bm{B}}\left(\bm{r}, t\right)&\coloneqq& \frac{1}{\lambda}\bm{B}\left(\frac{\bm{r}}{\lambda}, \frac{t}{\lambda}\right).
\end{eqnarray}

The scale transformations with $\lambda=1$ and $\tau\ne 1$ considered at the end of the previous section 
are incompatible with Maxwell's equations.
However, in general, this does not cause a problem for linacs.
This is because in the case of a drift tube linac (DTL),
it is sufficient to consider electromagnetic fields near the beam axis
to calculate the ion trajectory, and therefore, $\bm{E}$ can be assumed to be the rf electric field and 
$\bm{B}$ the static magnetic field. 
In other words, this pair of fields does not satisfy Maxwell's equations.
Moreover, in radio-frequency quadrupole (RFQ) linacs, ion trajectories are obtained by considering the rf electric field between the vanes, 
and the rf magnetic field is not included in the equation of motion.
In other words, a scale transformation with $\lambda=1$ and $\tau\ne 1$ is approximately correct.

The scale transformations (\ref{eq:scalex})--(\ref{eq:scaleb}) considered above are interesting in the following sense. 
First, the speed of the ions is kept invariant. 
Second, the gap voltage in the case of DTL 
and the inter-vane voltage in the case of RFQ are invariant, as we scale everything including cavity boundaries.
Therefore, if we make a cavity 10 times larger than the original, for example, 
we can accelerate ions to the same speed as the original with the same voltage. 
As the strength of the electric field will be 1/10 and ions will take 10 times longer to fly, 
the impulse received by an ion 
will be the same as the original, and the trajectory will be scaled to 10 times the original trajectory.

The power dissipation caused by rf loss on the cavity surface is considered as follows:
the surface electric current becomes 1/10 owing to eq. (\ref{eq:scaleb});
therefore, the integral of the square of the surface current per unit area is 1/100 of the original.
It seems that the area of the cavity surface cancels out by a factor of 100, but as the skin depth becomes $\sqrt{10}$ times larger, 
the total loss will be reduced to $1/\sqrt{10}$ of the original, if the cavity is manufactured ideally.

\section{Scale transformation of Hill's equation}

We consider the scale transformation of the Hill's equation corresponding to the discussion in the previous section.
First, suppose that the function $x(s)$ satisfies the following equation for $K$ such that $K(s+L)=K(s)$, i.e.,
\begin{eqnarray}
\label{eq:hill}
\ddot{x}(s)+K(s)x(s)=0.
\end{eqnarray}
Then, what kind of $K$ would provide $X(s)$, which is defined by $X(s)\coloneqq \lambda x(s/\lambda)$, as a solution of Hill's equation?
Using, eq. (\ref{eq:hill}), we can observe that
\begin{eqnarray}
\ddot{X}(s)&=&\frac{1}{\lambda}\ddot{x}\left(\frac{s}{\lambda}\right)\nonumber\\
&=&-\frac{1}{\lambda}K\left(\frac{s}{\lambda}\right)x\left(\frac{s}{\lambda}\right)\nonumber\\
&=&-\frac{1}{\lambda^2}K\left(\frac{s}{\lambda}\right)X(s).
\end{eqnarray}
Therefore, if we define
\begin{eqnarray}
\label{eq:scalek}
\tilde{K}(s)\coloneqq\frac{1}{\lambda^2}K\left(\frac{s}{\lambda}\right),
\end{eqnarray}
$X(s)$ will satisfy Hill's equation with $\tilde{K}$,
\begin{eqnarray}
\ddot{X}(s)+\tilde{K}(s)X(s)=0,
\end{eqnarray}
such that
\begin{eqnarray}
\label{eq:newperiod}
\tilde{K}(s+\lambda L)=\tilde{K}(s).
\end{eqnarray}

Equation (\ref{eq:newperiod}) is consistent with the scale transformation of the electromagnetic fields considered in the previous section. 
The same is true for eq. (\ref{eq:scalek}).
If we consider a quadrupole electric field that is uniform along the beam axis, for example, 
we can set $|\bm{E}(\bm{r})|=b r$, where $b$ is a constant.
In this case, based on eq. (\ref{eq:scalee}), $|\tilde{\bm{E}}(\bm{r})|=b r/\lambda^2$, 
which implies that the focusing force becomes $1/\lambda^2$ times, as in eq. (\ref{eq:scalek}).

As the function $X(s)$ satisfies the following equation:
\begin{eqnarray}
\dot{X}(s)=\dot{x}\left(\frac{s}{\lambda}\right),
\end{eqnarray}
the distribution of particles in phase space does not spread further along the angular direction, and the emittance becomes $\lambda$ times larger.
Moreover, as the beta function is transformed by
\begin{eqnarray}
\tilde{\beta}(s)=\lambda\beta\left(\frac{s}{\lambda}\right),
\end{eqnarray}
ellipse parameters are transformed as follows:
\begin{eqnarray}
\tilde{\alpha}(s)&\coloneqq&-\frac{1}{2}\frac{d\tilde{\beta}}{ds}(s)\nonumber\\
&=&\alpha\left(\frac{s}{\lambda}\right),\\
\tilde{\gamma}(s)&\coloneqq&\frac{1+\tilde{\alpha}(s)^2}{\tilde{\beta}(s)}\nonumber\\
&=&\frac{1}{\lambda}\gamma\left(\frac{s}{\lambda}\right).
\end{eqnarray}

\section{\label{sec:sc}Scale transformation of space-charge electric field}

Next, consider the case where the motion of a fixed number of ions undergoes scale transformations (\ref{eq:scalex}) -- (\ref{eq:scaleb}).
If the charge density of the original ion distribution is expressed as $\rho(\bm{r}, t)$, 
it is assumed that the charge density after the transformation will be expressed as
\begin{eqnarray}
\label{eq:scalerho}
\hat{\rho}(\bm{r}, t)\coloneqq\frac{1}{\lambda^3}\rho\left(\frac{\bm{r}}{\lambda},\frac{t}{\lambda}\right),
\end{eqnarray}
as the number of ions does not change after the scale transformation.
On the other hand, 
let $\bm{E}_{sc}(\bm{r,t})$ denote the electric field representing the space charge effect caused by $\rho(\bm{r}, t)$,
i.e.,
\begin{eqnarray}
\epsilon_0\nabla\cdot\bm{E}_{sc}(\bm{r},t)=\rho(\bm{r},t).
\end{eqnarray}
Then, the space-charge electric field $\hat{\bm{E}}_{sc}(\bm{r},t)$ arising from $\hat{\rho}(\bm{r}, t)$ becomes
\begin{eqnarray}
\label{eq:scaleesc}
\hat{\bm{E}}_{sc}(\bm{r},t)=\frac{1}{\lambda^2}\bm{E}_{sc}\left(\frac{\bm{r}}{\lambda},\frac{t}{\lambda}\right),
\end{eqnarray}
because the following equation holds:
\begin{eqnarray}
\epsilon_0\nabla\cdot\hat{\bm{E}}_{sc}(\bm{r},t)&=&\frac{\epsilon_0}{\lambda^2}\nabla\cdot\bm{E}_{sc}\left(\frac{\bm{r}}{\lambda},\frac{t}{\lambda}\right)\nonumber\\
&=&\frac{1}{\lambda^3}\rho\left(\frac{\bm{r}}{\lambda},\frac{t}{\lambda}\right)\nonumber\\
&=&\hat{\rho}(\bm{r},t).
\end{eqnarray}

When the linear dimension of a linac is scaled by a factor of $\lambda$, the cell length is increased by a factor of $\lambda$.
In this case, based on eq. (\ref{eq:scaleesc}), it can be inferred that the momentum change (impulse) per unit cell caused by the space charge field is 
$1/\lambda$ times smaller if the number of ions is kept invariant, at least in the lowest order approximation.
This means that in a linac that is 10 times the size, if the number of ions is same, the effect of the space charge field will be 1/10.

\section{\label{sec:lag}Discussions}

We will reconsider what we have observed so far based on the Lagrangian 
of classical electrodynamics\footnote{For simplicity, we consider a single-particle system here, 
but the conclusion remains unchanged even for a multi-particle system. The discussion for the multi-particle case is given in the Appendix.}. 
The Lagrangian of non-relativistic electrodynamics is given by the sum of three terms as follows:
\begin{eqnarray}
\label{eq:l_tot}
L=L_m+L_{i1}+L_f,
\end{eqnarray}
where
\begin{eqnarray}
\label{eq:lm}
L_m&\coloneqq&\frac{m}{2}\dot{\bm{x}}(t)^2, \\
L_{i1}&\coloneqq&q\left(-\phi(\bm{x}(t),t)+\dot{\bm{x}}(t)\cdot\bm{A}(\bm{x}(t),t)\right),\\
L_f&\coloneqq&\int_{[\Omega]} d^3\bm{r}\left (\frac{\epsilon_0}{2}\bm{E}^2(\bm{r},t)-\frac{1}{2\mu_0}\bm{B}^2(\bm{r},t) \right).
\end{eqnarray}
The action integral is written as
\begin{eqnarray}
S=\int_{t_0}^{t_1} L dt,
\end{eqnarray}
and the dynamics of ions and fields are determined, so as to satisfy the principle of least action, $\delta S=0$.
To obtain the equations of motion under Lorentz force (\ref{eq:lorentz}) and the first set of Maxwell's equations
(Faraday's law and Gauss' law for magnetic field), 
we fix the fields and consider a variant with regard to the trajectory\cite{bib:ll1}.
To obtain the second set of Maxwell's equations
(Amp\`ere's law and Gauss' law for electric field), 
we fix the trajectory and consider a variant with regard to the quantity of the fields\cite{bib:ll2}.

Let us define the scale transformation with respect to time coordinate, 
spatial coordinate, trajectory, and electromagnetic fields as follows:
\begin{eqnarray}
\label{eq:t1}
\tilde{t}&\coloneqq&\frac{t}{\tau}, \\
\label{eq:r1}
\tilde{\bm{r}}&\coloneqq&\lambda\bm{r},\\
\label{eq:x1}
\bm{X}(t)&\coloneqq& \lambda\bm{x}(\tau t),\\
\label{eq:e1}
\tilde{\bm{E}}\left(\bm{r}, t\right)&\coloneqq& \lambda \tau^2\bm{E}\left(\frac{\bm{r}}{\lambda}, \tau t\right),\\
\label{eq:b1}
\tilde{\bm{B}}\left(\bm{r}, t\right)&\coloneqq& \tau\bm{B}\left(\frac{\bm{r}}{\lambda}, \tau t\right).
\end{eqnarray}
These are identical to the scale transformations considered in Section \ref{sec:scalemot}. 
Although not used in the following discussion, 
the action integral $S'$ written in the transformed variables is subject 
to the factor $1/\tau$ from eq. (\ref{eq:t1}).
In other words, as an integral operator, the following equation holds:
\begin{eqnarray}
\int_{t_0/\tau}^{t_1/\tau}dt'=\frac{1}{\tau}\int_{t_0}^{t_1}dt.
\end{eqnarray}

Based on the definition of electromagnetic potential,
\begin{eqnarray}
\bm{E}\left(\bm{r}, t\right)&=&-\nabla\phi\left(\bm{r}, t\right)-\frac{\partial\bm{A}}{\partial t}\left(\bm{r}, t\right),\\
\bm{B}\left(\bm{r}, t\right)&=&\nabla\times\bm{A}\left(\bm{r}, t\right),
\end{eqnarray}
it is easy to observe that the potential functions of the transformed electromagnetic fields are
\begin{eqnarray}
\label{eq:phi2}
\tilde{\phi}\left(\bm{r}, t\right)&\coloneqq&\lambda^2 \tau^2\phi\left(\frac{\bm{r}}{\lambda}, \tau t\right),\\
\label{eq:a2}
\tilde{\bm{A}}\left(\bm{r}, t\right)&\coloneqq&\lambda \tau\bm{A}\left(\frac{\bm{r}}{\lambda}, \tau t\right).
\end{eqnarray}

Next, consider the scale transformation of each term of the Lagrangian.
First, the ``matter'' term $L_m$ is transformed as 
\begin{eqnarray}
\label{eq:lm2}
\tilde{L}_m&\coloneqq&\frac{m}{2}\left(\frac{d\bm{X}\left(\tilde{t}\right)}{d\tilde{t}}\right)^2\nonumber\\
&=&\frac{m}{2}\left(\lambda\tau\frac{d\bm{x}\left(t\right)}{dt}\right)^2\nonumber\\
&=&\lambda^2 \tau^2 L_m.
\end{eqnarray}
Second, using eqs. (\ref{eq:phi2}) and (\ref{eq:a2}), the ``interaction'' term $L_{i1}$ is transformed as
\begin{eqnarray}
\label{eq:li12}
\tilde{L}_{i1}&\coloneqq&q\left(-\tilde{\phi}\left(\bm{X}\left(\tilde{t}\right), 
\tilde{t}\right)+\frac{d\bm{X}\left(\tilde{t}\right)}{d\tilde{t}}\cdot\tilde{\bm{A}}\left(\bm{X}\left(\tilde{t}\right), \tilde{t}\right)\right)\nonumber\\
&=&\lambda^2 \tau^2 L_{i1}.
\end{eqnarray}
Thus, $L_m$ and $L_{i1}$ are transformed based on the same proportionality factor.
This is consistent with what we observed in the first half of Section \ref{sec:scaleem}.

On the other hand, the ``field'' term $L_f$ is transformed into
\begin{eqnarray}
\tilde{L}_f&\coloneqq&\int_{[\lambda^3\Omega]} d^3\tilde{\bm{r}}\left (\frac{\epsilon_0}{2}\tilde{\bm{E}}^2(\tilde{\bm{r}},\tilde{t})-\frac{1}{2\mu_0}\tilde{\bm{B}}^2(\tilde{\bm{r}},\tilde{t}) \right)\nonumber\\
&=&\lambda^3\int_{[\Omega]} d^3\bm{r}\left (\frac{\epsilon_0}{2}\lambda^2\tau^4\bm{E}^2(\bm{r},t)-\frac{1}{2\mu_0}\tau^2\bm{B}^2(\bm{r},t) \right).\nonumber\\
&&
\label{eq:lf0}
\end{eqnarray}
In order for this to be proportional to $L_f$, the constraint (\ref{eq:kousoku}) is sufficient, 
as we observed in the second half of Section \ref{sec:scaleem}.
Under this constraint, equation (\ref{eq:lf0}) reduces to
\begin{eqnarray}
\label{eq:lf2}
\tilde{L}_f=\lambda L_f,
\end{eqnarray}
and the source-free Maxwell's equations become scale invariant.

However, the proportionality factor appearing in the right-hand side of eq. (\ref{eq:lf2}) is different from those of eqs.
(\ref{eq:lm2}) and (\ref{eq:li12}),  
which means that Maxwell's equations are not scale invariant when the source terms are included.
This can be seen from the fact that the transformation factor of 
the space-charge electric field (\ref{eq:scaleesc}) is different from the factor of the electric field (\ref{eq:scalee}).

If we rewrite the transformations (\ref{eq:t1})--(\ref{eq:b1}) under the constraint $\lambda\tau=1$, 
we have the following definitions, which partially overlap with the previous ones:
\begin{eqnarray}
\label{eq:t2}
\tilde{t}&\coloneqq&\lambda\tau, \\
\label{eq:r2}
\tilde{\bm{r}}&\coloneqq&\lambda\bm{r},\\
\label{eq:x2}
\bm{X}(t)&\coloneqq& \lambda\bm{x}\left(\frac{t}{\lambda}\right),\\
\label{eq:e2}
\tilde{\bm{E}}\left(\bm{r}, t\right)&\coloneqq& \frac{1}{\lambda}\bm{E}\left(\frac{\bm{r}}{\lambda}, \frac{t}{\lambda}\right),\\
\label{eq:b2}
\tilde{\bm{B}}\left(\bm{r}, t\right)&\coloneqq& \frac{1}{\lambda}\bm{B}\left(\frac{\bm{r}}{\lambda}, \frac{t}{\lambda}\right).
\end{eqnarray}

We will further consider the scale invariance of Maxwell's equation under the constraints $\lambda\tau=1$.
Let us first rewrite the interaction term based on the charge density and current density as follows:
\begin{eqnarray}
L_{i2}\coloneqq\int_{[\Omega]} d^3\bm{r}\left(-\rho(\bm{r},t)\phi(\bm{r},t)+\bm{j}(\bm{r},t)\cdot\bm{A}(\bm{r},t)\right).\nonumber\\
\end{eqnarray}
Then, instead of eq. (\ref{eq:scalerho}), we define the scale transformation of the source terms as follows:
\begin{eqnarray}
\label{eq:rho2}
\tilde{\rho}(\bm{r}, t)&\coloneqq&\frac{1}{\lambda^2}\rho\left(\frac{\bm{r}}{\lambda},\frac{t}{\lambda}\right),\\
\label{eq:i2}
\tilde{\bm{j}}(\bm{r}, t)&\coloneqq&\frac{1}{\lambda^2}\bm{j}\left(\frac{\bm{r}}{\lambda},\frac{t}{\lambda}\right).
\end{eqnarray}
This corresponds to an operation that 
{\bf increases the overall charge of the ions by a factor of $\lambda$ simultaneously with the scale transformation}.
With these definitions, the interaction term $L_{i2}$ is transformed as 
\begin{eqnarray}
\label{eq:li22}
\tilde{L}_{i2}&\coloneqq&\int_{[\lambda^3\Omega]} d^3\tilde{\bm{r}}\left(-\tilde{\rho}(\tilde{\bm{r}},\tilde{t})\tilde{\phi}(\tilde{\bm{r}},\tilde{t})+\tilde{\bm{j}}(\tilde{\bm{r}},\tilde{t})\cdot\tilde{\bm{A}}(\tilde{\bm{r}},\tilde{t})\right)\nonumber\\
&=&\lambda\int_{[\Omega]} d^3\bm{r}\left(-\rho(\bm{r},t)\phi(\bm{r},t)+\bm{j}(\bm{r},t)\cdot\bm{A}(\bm{r},t)\right)\nonumber\\
&=&\lambda L_{i2}.
\end{eqnarray}
Now the proportionality factor in the right hand side is the same as that of eq. (\ref{eq:lf2}).
This means that the whole set of Maxwell's equations becomes scale invariant even with the source terms.
It can also be directly proved that the transformed electromagnetic field (\ref{eq:e2})--(\ref{eq:b2}),
together with the new source terms(\ref{eq:rho2})--(\ref{eq:i2}), satisfy Maxwell's equations.

Finally, we consider scale invariance, including the matter term $L_m$.
When we {\bf transform the mass and charge} as
\begin{eqnarray}
\label{eq:m2}
\tilde{m}&\coloneqq& \lambda m,\\
\label{eq:c2}
\tilde{q}&\coloneqq& \lambda q,
\end{eqnarray}
matter term $L_m$ and interaction term $L_{i1}$ are transformed by multiplying the common factor $\lambda$, based on eqs. (\ref{eq:lm2}) and (\ref{eq:li12}), i.e., 
\begin{eqnarray}
\label{eq:lm_tr}
\tilde{L}_m&=&\lambda L_m,\\
\tilde{L}_{i1}&=&\lambda L_{i1}.
\end{eqnarray}
As equation (\ref{eq:c2}) is consistent with eqs. (\ref{eq:rho2})--(\ref{eq:i2}), equation (\ref{eq:li22}) also holds.
Combining these results with eq. (\ref{eq:lf2}), 
we can see that all the terms in Lagrangian are transformed by multiplying the common factor $\lambda$.
This implies that {\bf all the equations of electrodynamics become scale invariant},
including the equation of motion under Lorentz force and Maxwell's equation with source terms.
This is a condition of mechanical similarity in classical electrodynamics.
The charge-to-mass ratio $q/m$ is invariant, but the electric charge and current are $\lambda$ times larger; 
therefore, the equation of motion is scaled to be consistent with the electromagnetic fields including the space charge fields.
This is consistent with the discussion of the space charge field in Section \ref{sec:sc}.
Moreover, Maxwell's equations are satisfied with transformed electromagnetic fields and source terms.

Based on the aforementioned considerations, let us consider a specific 
case where a cavity of an RFQ linac for deuterons is enlarged by a linear factor of 10. 
Suppose that the original RFQ was designed to accelerate deuterons at 100 mA\cite{bib:com18}. 
Then, based on the aforementioned considerations, the enlarged RFQ can accelerate the same number of $^{20}$Ne$^{10+}$ ions
up to the same energy, if the same value of the inter-vane voltage as the original is applied to the enlarged cavity. 
The beam current in this case is 1 A.
The trajectories will be scaled by exactly 10 times, even in the presence of the space charge fields. 

In a very crude approximation, one might think that one $^{20}$Ne$^{10+}$ ion could be replaced by ten deuterons.
If this replacement holds, the enlarged RFQ considered above may accelerate deuterons with a beam current of 1 A.
It will be interesting to examine the possibility of the replacement by computer simulations, and this is a subject for future studies.
If successful, this ``Big RFQ'' can be an alternative candidate for a high current linac for nuclear transmutation\cite{bib:oku19}.

\section{\label{sec:summary}Summary}

The set of scale transformations (\ref{eq:t2})--(\ref{eq:b2}), (\ref{eq:rho2})--(\ref{eq:i2}), and (\ref{eq:m2})--(\ref{eq:e2})
makes all the equations of electrodynamics invariant, i.e., 
the equation of motion under Lorentz force and Maxwell equations with source terms. 
The invariance is supported by the mechanical similarity that appears in the Lagrangian of classical electrodynamics. 
This scale transformation may help in the initial design of high-current accelerators, although further simulation study is required.

\appendix*
\section{Multi-particle case}
We consider a system consisting of $N$
 charged particles, where the mass, charge, and trajectory of the $i$-th particles are denoted by $m_i,~q_i,$ and $\bm{x}_i(t)$, respectively.
The charge density and current density of this system are given by the following expressions.
\begin{eqnarray}
\label{eq:rho_m}
\rho'(\bm{r}, t)&\coloneqq&\sum_{i=1}^N q_i\delta_3(\bm{r}-\bm{x}_i(t)),\\
\label{eq:i_m}
\bm{j}'(\bm{r}, t)&\coloneqq&\sum_{i=1}^N q_i\dot{\bm{x}}(t)\delta_3(\bm{r}-\bm{x}_i(t)).
\end{eqnarray}
where $\delta_3(\bm{r})$ is defined as the product $\delta(x)\delta(y)\delta(z)$.
It should also be noted that $\rho'$ and $\bm{j}'$ satisfy the continuity equation.

Performing the scale transformation of $\rho'$ according to eq. (\ref{eq:rho2}) yields the following result.
\begin{eqnarray}
\label{eq:rho3}
\tilde{\rho'}(\bm{r}, t)&\coloneqq&\frac{1}{\lambda^2}\rho'\left(\frac{\bm{r}}{\lambda},\frac{t}{\lambda}\right)\nonumber\\
&=&\frac{1}{\lambda^2}\sum_{i=1}^N q_i\delta_3\left(\frac{\bm{r}}{\lambda}-\bm{x}_i\left(\frac{t}{\lambda}\right)\right)\nonumber\\
&=&\lambda\sum_{i=1}^N q_i\delta_3\left(\bm{r}-\lambda\bm{x}_i\left(\frac{t}{\lambda}\right)\right)\nonumber\\
&=&\lambda\sum_{i=1}^N q_i\delta_3\left(\bm{r}-\bm{X}_i(t)\right).
\end{eqnarray}
This corresponds to a situation in which each particle is located at $\bm{X}_i(t)$ and carries a charge of $\lambda q_i$.
Similarly, following eq. (\ref{eq:i2}) gives the corresponding result for the current density:
\begin{eqnarray}
\label{eq:i3}
\tilde{\bm{j}'}(\bm{r}, t)=&\coloneqq&\frac{1}{\lambda^2}\bm{j}'\left(\frac{\bm{r}}{\lambda},\frac{t}{\lambda}\right)\nonumber\\
&=&\lambda\sum_{i=1}^N q_i\dot{\bm{X}}(t)\delta_3\left(\bm{r}-\bm{X}_i(t)\right).
\end{eqnarray}
Since $\tilde{\rho'}$ and $\tilde{\bm{j}'}$ are constructed from $\rho'$ and $\bm{j}'$ according to eqs. (\ref{eq:rho2}) and (\ref{eq:i2}),
the interaction Lagrangian constructed from them, $\tilde{L}'_{i2}$ and $L'_{i2}$,
immediately satisfy eq. (\ref{eq:li22}), that is,
\begin{eqnarray}
\label{eq:li22_m}
\tilde{L}'_{i2}=\lambda L'_{i2}.
\end{eqnarray}

On the other hand, for the kinetic term, instead of eq. (\ref{eq:lm}), we have a sum over the kinetic energies of the $N$ particles.
\begin{eqnarray}
L'_m&\coloneqq&\sum_{i=1}^N\frac{m_i}{2}\dot{\bm{x}}_i(t)^2.
\end{eqnarray}
This term transforms under the scale transformation and the mass transformation 
$m_i\rightarrow \lambda m_i$
 in the same way as eq. (\ref{eq:lm_tr}), namely, 
$L'_m\rightarrow \lambda L'_m$.
Therefore, the total Lagrangian $L$ is multiplied by $\lambda$: $L\rightarrow \lambda L$.

\end{document}